\documentclass[prb,aps,twocolumn]{revtex4}
\usepackage{amsmath}
\usepackage{amsfonts}
\usepackage{amssymb}
\usepackage{graphicx}

\begin{document}

\title{Eigenfunction fractality and pseudogap state near
superconductor-insulator transition}
\author{M. V. Feigel'man$^1$, L. B. Ioffe$^{2,1}$, V. E. Kravtsov$^{3,1}$
and E. A. Yuzbashyan$^2$}
\affiliation{$^1$ L. D. Landau Institute for Theoretical Physics, \\
Kosygin str.2, Moscow 119334, Russia}
\affiliation{$^2$ Serin Physics Laboratory, Department of Physics and Astronomy,\\
Rutgers University, Piscataway, NJ 08854, USA}
\affiliation{$^3$ Abdus Salam International Center for Theoretical Physics, Trieste, Italy}
\date{\today}

\begin{abstract}
We develop a theory of a pseudogap state appearing near the
superconductor-insulator transition in strongly disordered metals with
attractive interaction. We show that such an interaction combined with the
fractal nature of the single particle wave functions near the mobility edge
leads to an anomalously large single particle gap in the superconducting
state near SI transition that persists and even increases in the insulating
state long after the superconductivity is destroyed. We give analytic
expressions for the value of the pseudogap in terms of the inverse
participation ratio of the corresponding localization problem.
\end{abstract}

\maketitle

\renewcommand{\L}{L_{\rm loc}}

Rapidly growing number of experiments~\cite%
{Shahar1992,Goldman1993,Kowal1994,Gantm96,Adams2001,Baturina2004,Shahar2004,
Baturina2003-6,Steiner2005} on various disordered superconductors show that
a novel phase often appears on the insulating side of the
superconductor-insulator transition. As the disorder strength is increased
the superconductivity is suppressed leading to a strange insulator
characterized by a large thermally assisted resistance with a small but hard
gap (Fig.~1). Further increasing the disorder one gets the usual variable
range hopping behavior. Experimentally, the phase diagram of disordered
superconductors is often explored by varying the applied magnetic field. On
the superconducting side of the transition a relatively small field destroys
the superconductivity resulting in a hard gap insulating state. At larger
magnetic fields the resistance and the gap drop~\cite%
{Gantm96,Adams2001,Baturina2004,Shahar2004,Baturina2003-6}. This is observed
only in a narrow window of disorder strengths, away from this window on a
superconducting side the application of magnetic field converts the
supercondcutor into a normal metal as usual.

It is tempting to explain these data by the formation of localized Cooper
pairs\cite{Gantm96,Gantmakher1998}. In this picture the superconductivity is
due to a fragile coherence between localized Cooper pairs, while the energy
to break the pair is much larger and remains finite even when the coherence
(and thus the superconductivity) is destroyed. The hypothesis of preformed
Cooper pairs is further confirmed by the behavior of these superconductors
at higher temperatures. On the insulating side of the transition in thick
(effectively 3D) films one observes~\cite{Shahar1992,Kowal1994} Arrenihus
temperature behavior of the resistivity, $R(T)\propto \exp (T_{I}/T)$, at
low temperatures. The experimental value of the activation energy, $T_{I}$,
is somewhat larger than the superconducting gap in less disordered samples
and grows with the disorder~\cite{Shahar1992,Kowal1994}. However, at higher
temperatures this behavior is replaced~\cite{Kowal1994} by Mott's variable
range hopping $R(T)\sim \exp (T_{M}/T)^{1/4}$. This can be understood if the
insulating pseudogap is due to preformed Cooper pairs with a relatively
large pairing energy $T_I$. 
\begin{figure}[th]
\label{PhaseDiagram}\includegraphics[width=2in]{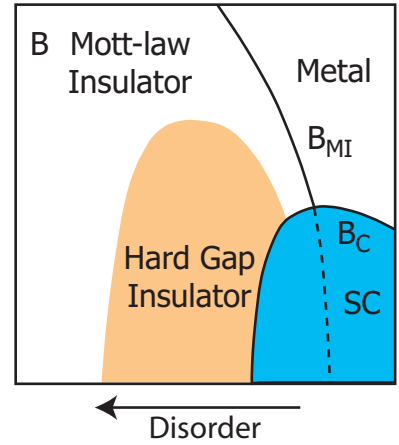}
\caption{Schematics of low-temperature phase diagram of disordered
superconductors in the vicinity of the superconductor-insulator transition
as a function of the magnetic field, $B$, and disorder strength. }
\end{figure}
\qquad

In this Letter we show that 
preformed Cooper pairs appear in the semi-microscopic three-dimensional
model that contains only low energy electrons with weak BCS-type attraction
and a strong random potential that leads to Anderson localization of
single-particle states.

In the presence of preformed Cooper pairs the parity effect should arise --
the ground state energy with even number of electrons is lower then that for
the closest odd number. The corresponding \textit{parity gap} $%
\Delta_{P}\sim T_{I}\sim \delta/\ln(\delta/\Delta)$ has been calculated by
Matveev and Larkin \cite{Matveev1997} for small superconducting grains. Here 
$\delta=1/(\nu_{0}L^{3})$ is the mean level spacing in the grain,  $\Delta<<
\delta$ is the energy gap in the bulk superconductor, and $\nu_{0}$ and $L^3$
are the density of orbital states and the volume of the grain, respectively.
In this Letter we argue that the result of Ref.\cite{Matveev1997} can be
generalized to \textit{bulk} Anderson insulators. In this case, $L$ is
replaced by the localization radius $L_{\mathrm{loc}}$, so that $%
\delta\rightarrow\delta_{L}=1/(\nu_{0}L_{\mathrm{loc}}^{3})$, and the BSC $%
\Delta$ is replaced by the superconducting gap $\Delta_{\mathrm{crit}}$ at
the Anderson transition point. The fractal nature of
near-critical wavefunctions characterized by the fractal dimension $D_{2}<3$
determines the reduction factor $(\delta_{L}/\Delta_{\mathrm{crit}%
})^{1-D_{2}/3}\gg 1$ that replaces $\ln(\delta/\Delta)$ in the
Matveev-Larkin formula.

We consider two different regimes. 
In more disordered materials where $\delta _{L}\gg \Delta_{\mathrm{crit}} $
the \thinspace\ Cooper instability and superconductive long-range order
disappear. However, the attraction between electrons persists as long as $%
\delta _{L}$ remains smaller than the Debye frequency $\omega _{D}$ and
results in the "local" pairing of electrons with opposite spins occupying
the same localized state. 
We show below that in this regime the \emph{hard-gap insulator }is formed
whith properties similar to those observed in\cite{Shahar1992,Kowal1994}. We
further argue that in the less localized regime, when $\delta _{L}\leq
\Delta_{\mathrm{crit}}$, the unusual superconductive state with a \emph{%
pseudogap} is formed. The 
features of this state are: (i) single-electron excitation gap $\Delta _{1}$
is larger than superconductive gap $\Delta $ so that the ratio $\Delta
_{1}/T_{c}$ is anomalously high, and (ii) insulating trend in the $R(T)$
curves exists above $T_{c}$.

We assume that superconductivity is due to attraction between electrons that
originates at high energy scales $\sim \omega _{D}$ and that it is affected
by localization of electron wave functions only for a very large disorder $%
\delta _{L}\gtrsim \omega _{D}$. In a fermion system with weak attraction
one can leave only the pair interaction terms in the Hamiltonian leading to
the usual BCS model in the basis of localized electron states~\cite{Ma1985}: 
\begin{eqnarray}
H &=&\sum_{j\sigma }\epsilon _{j}c_{j\sigma }^{\dagger }c_{j\sigma }-\frac{%
\lambda }{\nu _{0}}\sum_{j,k}M_{jk}c_{j\uparrow }^{\dagger }c_{j\downarrow
}^{\dagger }c_{k\uparrow }c_{k\downarrow }\,,  \notag \\
\mathrm{where}\qquad M_{jk} &=&\int d\mathbf{r}\psi _{j}^{2}(\mathbf{r)}\psi
_{k}^{2}(\mathbf{r})\,,  \label{Ham}
\end{eqnarray}%
$\lambda $ is dimensionless Cooper coupling constant, $\epsilon _{j}$ is the
single-particle energy of the state $j$, and $c_{j\sigma }$ is the
corresponding electron operator for the spin projection $\sigma $.

Physical properties of the electron system are controlled by the electrons
near the Fermi level, so a very important implicit ingredient of the model (%
\ref{Ham}) is the statistics of matrix elements $M_{jk}$ between eigenstates
in the vicinity of Anderson mobility edge. The key feature of these
nearly-critical wavefunctions is their \emph{fractal} structure~\cite%
{MirlinReview2000} that shows in the anomalous scaling of diagonal matrix
elements $M_{jj}\equiv M_{j}$ ("inverse participation ratios", IPRs) with
localization length: typical IPR $\bar{M}\propto L_{loc}^{-D_{2}}$, where 
\emph{fractal dimension} $D_{2}<3$. Numerical studies \cite%
{Parshin1999,Mirlin2002} indicate that $D_{2}=1.30\pm 0.05$ for the standard
3D Anderson transition. 
The IPR distribution function $\mathcal{P}(M_{j})$ has been studied in \cite%
{Mirlin2002}. Scaling theory of localization predicts that near the mobility
edge $\mathcal{P}(M_{j})$ acquires a scale-invariant form and this is indeed
what was observed \cite{Mirlin2002}. The same data demonstrate that $%
\mathcal{P}(M_{j})$ decreases fast for atypicallly extended states, i.e. at $%
M_{j}/\bar{M}\ll 1$. This allows us to use the average value 
\begin{equation}
\bar{M}=L_{0}^{-3}(L_{loc}/L_{0})^{-D_{2}},  \label{M}
\end{equation}%
where $L_{0}$ is the short-scale cutoff length of the fractal behaviour. The
associated energy scale $E_{0}=1/(\nu _{0}L_{0}^{3})$ depends on the
microscopic details of the model of disorder and might be small compared to
Fermi-energy $E_{F}$. Localization length depends on Fermi-energy (in the
scaling region $L_{loc}\gg L_{0}$) as 
$L_{loc}\approx L_{0}(E_{0}/({E_{m}-E_{F}}))^{\nu } $, 
where $E_{m}$ is the position of the mobility edge and $\nu $ is the
localization length exponent.

Another important property of the nearly-critical eigenstates is their
strong correlation in energy and real space ~\cite{Chalker1990, KrMut1997,
MirlinReview2000} even in the limit of strong fractality $D_{2} << 3$.  It
results in the scaling dependence of the \emph{average} matrix elements $%
\overline{M_{jk}}$ on the energy difference $E_{j}-E_{k}=\omega $: 
\begin{equation}
\mathcal{V}\overline{M}_{jk}\equiv M(\omega )\approx \left\{ 
\begin{array}{rc}
(L_{loc}/L_{0})^{3\gamma }\quad \mathrm{at}\quad \omega \ll \delta _{L} & (a)
\\ 
(E_{0}/\omega )^{\gamma }\quad \mathrm{at}\quad \delta _{L}\ll \omega \leq
E_{0} & (b)%
\end{array}%
\right.  \label{MM}
\end{equation}%
where $\gamma =1-D_{2}/3$ and $\mathcal{V}$ is the total system's volume.
Note that in the critical region $M(\omega)\gg 1$ in contrast both to a
metal and to a deep insulator ($L_{loc} \sim L_{0}$), where $M(\omega)
\approx 1$.

We begin with the insulating region $\Delta \ll \delta _{L}\ll \omega _{D}$
where Cooper interaction can be treated perturbatively. In the first order
of the perturbation theory we take into account only diagonal terms $j=k$ of
the interaction similar to the case of ultrasmall grain~\cite{Matveev1997}. 
Then the energy (counted from $E_{F}$) required to break a bound pair of
electrons siting in the $j$-th orbital state is $2\Delta _{P}^{(j)}=\frac{%
\lambda }{\nu _{0}}M_{j}$. Typical value of this "parity gap" (cf.~\cite%
{Matveev1997}) scales then as 
\begin{equation}
\Delta _{P}=\frac{{\lambda }}{2}E_{0}\left( \frac{L_{\mathrm{0}}}{L_{\mathrm{%
loc}}}\right) ^{D_{2}}\propto (E_{m}-E_{F})^{\nu D_{2}}  \label{DeltaP3}
\end{equation}%
Neglecting fluctuations in the local values of $\Delta _{P}^{(j)}$, one
finds that all states occupied by single electrons are shifted up by the
amount $\Delta _{P}$ that leads to the electron DoS $\tilde{\nu}(\varepsilon
)=\nu _{0}\theta (\varepsilon -\Delta _{P})$. In fact, values of $\Delta
_{P}^{(j)} $ differ for different localized states, and the average density
of states $\nu (\varepsilon )$ in a large sample is determined by the IPR
distribution $\mathcal{P}(M)$: 
\begin{equation}
\nu (\varepsilon )=\nu _{0}\int_{0}^{2\varepsilon \nu _{0}/\lambda }\mathcal{%
P}(M)dM  \label{DoS}
\end{equation}%
As mentioned above, numerical data for $\mathcal{P}(M)$ indicate its very
fast decrease at $M/\bar{M}\rightarrow 0$. Thus the DoS shape (\ref{DoS}) is
not far from a rectangular sharp gap, with the gap value given by Eq.~(\ref%
{DeltaP3}). We emphasize that i) parity gap $\Delta _{P}$ is much larger
than level spacing $\delta _{L}$ at $L_{loc}/L_{0}>(2/\lambda )^{\frac{1}{%
3-D_{2}}}$, and ii) the DoS (\ref{DoS}) does not contain any "coherence
peak" above the gap (cf. Ref.~\onlinecite{Ghosal2001}).

We associate~\cite{comment1} the spectral gap $\Delta _{P}$ with the
measured~\cite{Shahar1992,Kowal1994,Gantm96} activation energy $T_{I}$. The
external parameter $(E_m-E_F)$ representing the disorder strength in Eq.~(%
\ref{DeltaP3}) can be replaced with an experimentally more accessible
parameter $(\sigma_c-\sigma)\propto (E_m-E_F)$. Here $\sigma$ is the high
temperature conductivity and $\sigma_c$ is the value of the conductivity
where the parity gap $\Delta_P$ first develops. We obtain 
\begin{equation}
T_I=A(1-\sigma/\sigma_c)^{\nu D_2},  \label{ti}
\end{equation}
where $A$ is conductivity-independent. This equation predicts a moderate
increase of $T_{I}$ with disorder strength in agreement with the
experimental data \cite{Shahar1992}, see Fig. 2. 
\begin{figure}[th]
\includegraphics[width=2in]{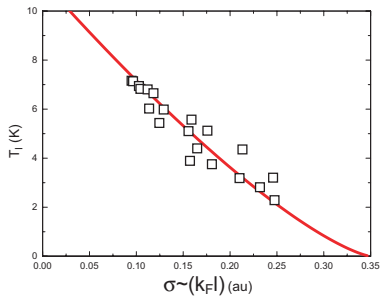}
\caption{Experimental values of the gap from Ref.\onlinecite{Shahar1992}, $%
T_{I}$ (boxes) and a fit to the equation (\protect\ref{ti}) with $\protect%
\nu =1$, $D_{2}=1.3$. The only fitting parameter was the constant $A=(%
\protect\lambda/2)E_{0}$; the data points of Ref.[1] correspond to $%
E_{0}\approx 100 K$ at $\protect\lambda\approx 0.2$ extracted from the BCS
value of $T_{c} \approx 3 K$ for less disordered samples~\protect\cite%
{Shahar1992} and $\protect\omega_D \approx 500 K$. The value of $\protect%
\sigma _{c}$ was determined from high $T$ data. Application of scaling
formulas to the data shown here is justified by the large value of the
localization length of the most disordered sample shown in this plot: $%
L_{loc}^{max} > 30 \mathring{A}$ which was deduced from the Mott temperature
characterizing the resistivity of similar samples at intermediate
temperatures. \protect\cite{Kowal1994} }
\label{DataFit}
\end{figure}

We now turn to the parameter region $\delta _{L}\ll \Delta $ where one
expects a global superconductive coherence to exist at low enough $T$.
Indeed, in this regime a given localized single-particle state typically
overlaps in a real space with a large number $\sim \Delta/\delta_L$ of
eigenstates $\psi _{j}$  in the same energy strip $|\epsilon_{j}|\leq \Delta$%
.  It is natural to expect that in this case the mean-field approximation
can provide some guidance, while not necessarily being quantitatively
accurate (e.g. because fluctuations of the overlap matrix elements $M_{jk}$
are large). To test the validity of the mean-field, we compared its
prediction for $T_{c}(\delta _{L})$ (see below) with the transition
temperature that was found numerically by computing the first terms of the
virial expansion applied to (\ref{Ham}) with $M_{ij}$ determined by exact
diagonalization. Reasonably good agreement was found \cite{Kravtsov06}.

To proceed with the mean-field analysis, we introduce averaged
energy-dependent pairing amplitudes $F(\epsilon _{j})=\overline{\langle
c_{j\uparrow }c_{j\downarrow }\rangle }$ and the gap function $\Delta
(\epsilon )=\lambda \int d\epsilon _{1}M(\epsilon -\epsilon _{1})F(\epsilon
_{1})$. Following standard steps, we decouple interaction term in the
Hamiltonian (\ref{Ham}) via the gap function $\Delta (\epsilon )$, calculate
anomalous averages $F(\epsilon )$, and arrive at the modified BSC gap
equation in the form 
\begin{equation}
\Delta (\epsilon )=\frac{\lambda }{2}\int_{-\infty }^{+\infty }d\epsilon _{1}%
\frac{M(\epsilon -\epsilon _{1})\Delta (\epsilon _{1})}{\sqrt{\epsilon
_{1}^{2}+\Delta ^{2}(\epsilon _{1})}}\tanh \frac{\sqrt{\epsilon
_{1}^{2}+\Delta ^{2}(\epsilon _{1})}}{2T}  \label{DeltaNew}
\end{equation}%
Gap function $\Delta (\epsilon )$ obeying Eq.(\ref{DeltaNew}) is an even
function of $\epsilon $ with the maximum value $\Delta (\epsilon =0)\equiv
\Delta_0 $.

Superconducting transition temperature $T_{c}$ is determined by
linearization of Eq.(\ref{DeltaNew}) with respect to $\Delta (\epsilon )$. 
Due to power-law decrease of $M(\omega )$ at large arguments, the integral
in Eq.(\ref{DeltaNew}) converges and is dominated by $\omega\sim T$
("infrared superconductivity") so no upper cutoff is needed, contrary to
usual BSC problem. When Fermi level is very close to the Anderson mobility
edge $E_{m}$ and level spacing $\delta _{L}\sim
E_{0}((E_{m}-E_{F})/E_{0})^{3\nu }$ is negligibly small, one can use for $%
M(\omega )$ the Eq.(\ref{MM}b). Then the critical temperature is given by 
\begin{equation}
T_{c}^{0}(\lambda ,\gamma )=E_{0}\lambda ^{1/\gamma }C(\gamma )  \label{Tc}
\end{equation}%
where dimensionless function $C(\gamma )$ can be computed numerically. At
small $\lambda$ the value given by Eq.(\ref{Tc}) exceeds the BCS value $%
T_{BSC}\sim \omega_{D}e^{-1/\lambda}$. This may lead to a maximum in $T_{c}$
near the critical disorder. The zero-temperature energy gap $\Delta _{T=0}$
in the same limit $\delta _{L}\rightarrow 0$ is given by Eq.(\ref{Tc}) with $%
C(\gamma )$ replaced by another function $D(\gamma )$. We plot $D(\gamma )$
and $2D(\gamma)/C(\gamma)$ in Figure 3.

\begin{figure}[th]
\includegraphics[width=2.5in]{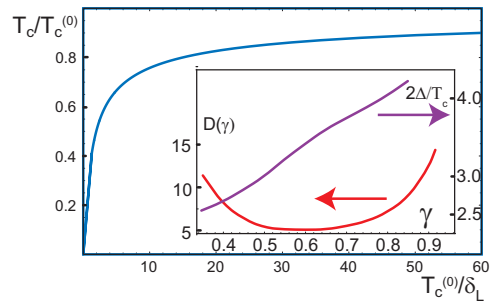}
\caption{Suppression of $T_c$ as a function of level spacing within
localization volume. The insert shows $\protect\gamma$ -dependence of the
dimensionless gap and $2 \Delta(0)/T_c$ for $\protect\delta_L=0$.}
\label{TcFigure}
\end{figure}
Using Eq.(\ref{Tc}) we eliminate the interaction constant $\lambda$ and the
cut-off parameter $E_{0}$ from Eq.(\ref{DeltaP3}) and arrive at 
\begin{equation}  \label{DeltaPfin}
\Delta_{P} = \frac1{2 (C(\gamma))^\gamma} \cdot \delta_{L}
\,\left(T_{c}^{0}/\delta_L \right)^{\gamma}
\end{equation}
This formula (applied at $\delta_L \gg T_c^0$) generalizes the results of
Ref.\cite{Matveev1997} to bulk strongly disordered superconductors. Note
that in contrast to Ref.\cite{Matveev1997} here the reduction of $\Delta_{P}$
compared to $\delta_{L}$ is not due to the renormalization of attractive
interaction~\cite{absense}, but to the enhancement of the matrix elements $%
M_{j}$ due to fractality.

To study the effect of nonzero level spacing $\delta _{L}$ upon $T_{c}$, we
approximate $M(\omega )$ by a simple interpolation formula $M(\omega
)=E_{0}^{\gamma }(\omega ^{2}+\delta _{L}^{2})^{-\frac{\gamma }{2}}$ and
solve the linearized version of Eq.(\ref{DeltaNew}) for $T_{c}(\delta _{L})$
numerically. Since $M(\omega )$ is a uniform function of $T$ and $\delta _{L}
$, while the coupling constant $\lambda $ and $E_0$ enter Eq.~(\ref{DeltaNew}%
) only in a combination $E_0\lambda^{1/\gamma}$, it is possible to present
the dependence $T_{c}(\delta _{L}) $ in the form 
\begin{equation}
T_{c}(\delta _{L})=T_{c}^{0}\mathcal{T}_{\gamma }\left( \frac{T_{c}^{0}}{%
\delta _{L}}\right)  \label{scalingTc}
\end{equation}%
where $T_{c}^{0}$ is defined in Eq.(\ref{Tc}) and scaling function $\mathcal{%
T}_{\gamma }(x)$ does not depend on $\lambda $. This universal function was
found numerically for $\gamma =0.57$ \thinspace\ 
(corresponding to $D_{2}=1.3 $ for 3D Anderson transition), the result is plotted in Fig.3.
Function $\mathcal{T}_{\gamma }(x)$ approaches very slowly its asymptotic
value 1 at large $x$. The actual $T_{c}$ is suppressed as compared to the
mean-field result  due to fluctuations. We expect it still has the scaling
form (\ref{scalingTc}) but with a different scaling function that vanishes
at $x\leq x_{c}\sim 1$.

Although gap equation (\ref{DeltaNew}) is similar to the conventional one,
the real-space properties of the state that it describes are unusual. 
The local pairing amplitude $F(\mathbf{r})=\sum_{j}\langle c_{j\uparrow
}c_{j\downarrow }\rangle \psi _{j}^{2}(\mathbf{r})$ is extremely
inhomogeneous in space, populating only small fraction $\propto \left[ {%
T_{c}(\delta _{L})}/{E_{0}}\right] ^{\gamma }$ of the total volume~\cite%
{fraction}. Diamagnetic response of such a superconductor differs strongly
from that of usual "dirty-limit" materials with uniform $|F(\mathbf{r})|$,
but reminds that of weakly-coupled Josephson junctions arrays.
Qualitatively, we expect (i) extremely weak Meissner effect and considerably
stronger linear shielding effect, and 
(ii) superconductor $\rightarrow $ gaped insulator $\rightarrow $ Mott
insulator sequence of transitions upon magnetic field increase.

We now discuss the effect of "local pairing" considered previously (see Eq.(%
\ref{DeltaP3})) in insulating state. We have seen that in insulating state
when $\delta _{L}\gg \Delta $ \textit{single-particle} excitation (carrying
spin $\frac{1}{2}$) have a gap $\Delta _{1}=\Delta _{P}$. On the other hand,
excitations that involve only hopping of paired electrons between localized
levels and do not involve breaking pairs are gapless, i.e. the \textit{%
energy gap} $\Delta $ vanishes. On the other hand, in superconducting state
within mean-field we have $\Delta _{1}=\Delta +\Delta _{P}$, while the gap
for pair excitations\cite{Anderson1958} (without pair breaking) is $\Delta
_{2}=2\Delta $. When $L_{loc}/L_{0}\rightarrow \infty $, the parity gap $%
\Delta _{P}$ becomes much larger than $\delta _{L}$. Therefore, we expect
that there is a regime $\Delta _{P}\gtrsim \Delta $ in the superconducting
state where the \textit{spin gap} $\Delta _{1}$ is larger than the energy
gap.

In the derivation of the gap equation (\ref{DeltaNew}) we neglected the
effects of the parity gap which is correct for  $\Delta _{P}\ll T$. A large
parity gap pushes up all unpaired states with respect to the paired ones
which modifies the thermal distribution factor in the Eq.(\ref{DeltaNew}).
In the limit of $\Delta _{P}\gg T$  the contribution of unpaired electrons
vanishes and the problem becomes equivalent to the pseudospin model \cite%
{Anderson1958}. This effectively doubles the value of the gap and thus
replaces $2T\rightarrow T$ in the $\tanh ()$ argument in (\ref{DeltaNew}).
For a general $\Delta _{P}\sim T$ the crossover formula between these two
regimes can be obtained in the mean-field approximation. The result is (for
derivation cf.~\cite{Kravtsov06} ) that $\tanh (...)$ in Eq.(\ref{DeltaNew})
should be replaced by 
\begin{equation}
\mathcal{F}(\epsilon _{1},T)=\frac{\sinh B}{\cosh B+e^{-\Delta _{P}/T}}\,,
\label{newdist}
\end{equation}
where $B=\sqrt{(\tilde{\epsilon}_{1}^{2}+\Delta ^{2}(\epsilon _{1})}/T$, and 
$\tilde{\epsilon}_{1}=\epsilon _{1}-\Delta _{P}$. 

Note that most experiments probing the superconducting gap, such as
tunnelling conductance, optical conductivity or NMR, measure the spin gap $%
\Delta _{1}$. One thus expects that these data would show anomalously large
(compared to BCS value of $1.76$) ratio of spectral gap to transition
temperature near S-I transition. An additional suppression of $T_{c}$ in
comparison with $\Delta _{P}$ can be due to electron-electron interaction in
the density channel (irrespectively of the interaction sign) that is not
included into the model (\ref{Ham}). 

An anomalously large ratio $\Delta_{1}/T_{c}$ leads to the insulating trend
of the resistivity versus temperature behavior in the intermediate
temperature range $T_{c}<T\leq\Delta _{1}$. This was observed in strongly
disordered superconductors, see e.g.~\cite{Baturina2003-6}, and is 
well-known as a \textit{pseudogap} phenomenon in underdoped cuprates~\cite%
{cuprates}. The quantitative similarity between $R(T,B)$ behavior in InO$_{x}
$ films and underdoped cuprates~\cite{Steiner2004,Sun2005} allows one to
speculate that the pseudogap in underdoped cuprates might also be related to
pairing of electrons on localized states. The important difference of the
cuprates is the d-wave symmetry of the pairing.

The effect of magnetic field on the fractal superconductor deserves a
separate study. Here we only remark that the available magnetoresistance
data are not consistent with a conventional picture of weakly coupled
superconducting grains but point towards fractal superconductivity. 
Typically,~\cite{Gantm96,Adams2001,Baturina2004,Shahar2004,Steiner2005}, the
magnetoresistance is negative at high fields and persists up to very high
field values. Consider, for instance, the magnetoresistance data on
amorphous InO$_{x}$ shown in Fig.1 of Ref.~\onlinecite{Gantm96} (the samples
were of the same type and origin as those studied in~\cite%
{Shahar1992,Kowal1994}). Activation energy for this sample is $T_{I}\approx
15\,\mathrm{K}$ (as determined in the temperature range $1.3-5$ Kelvin);
application of 8 Tesla field at the temperature of 1.58 K leads to negative
magnetoresistance (MR) $R(H=8\mathrm{T})/R(0)\approx 0.5$. Interpreting this
effect as being due to magnetic-field suppression of the pairing gap in
small grains, we find $T_{I}-T_{I}(H=8\mathrm{T})\approx 0.7\,\mathrm{K}$,
which is about 5\% of $T_{I}$ only. Interpolating this dependence we
estimate that the field necessary to destroy completely superconductivity in
each grain is about $H_{cg}^{\exp }\sim 50-80\mathrm{Tesla}$ \cite{St2}
which is much larger than one expects for realistic grain sizes. Indeed,
orbital critical magnetic field for a small (radius $R<\xi $, where 
$\xi =\sqrt{\hbar D/\Delta }$ is the coherence length) superconductive grain is
equal~\cite{Larkin1965,Bel2000} to $H_{cg}^{est}\approx \frac{1000\,\mathrm{T
}}{R\xi }$ where $R$ and $\xi $ are measured in nanometers. Using a typical
diffusion constant for a poor metal $D\approx 1\mathrm{cm}^{2}/\mathrm{s}$
and allowing for a very high gap value $\Delta =10\,\mathrm{K}$, we find $%
\xi =8.5\,\mathrm{nm}$. Together with the lowest bound for the grain radius $%
R=6\,\mathrm{nm}$ (determined via the condition $\delta =(4\nu
_{0}R^{3})^{-1}<\Delta $) it leads to $H_{cg}^{est}\leq 20\,\mathrm{Tesla}$
which is still much smaller than $H_{cg}^{\exp }$ above. These estimates did
not take into account the spin effect of magnetic field that would further
decrease the value of $H_{cg}^{est}$. This discrepancy between the
experimental value of the characteristic magnetic field, $H_{cg}^{\exp }$,
and its estimate, $H_{cg}^{est}$ can be accounted for by the fractal nature
of the paired states because the effect of magnetic field on the local
pairing gets much weaker due to  low value of fractal dimension $%
D_{2}=1.3$ of these states that indicates "almost one-dimensional" nature of
these eigenstates.

In conclusion, weak Anderson insulators with Cooper attraction are shown to
possess hard insulating gap whose magnitude is determined by the IPR
statistics near the mobility edge. Although this gap is due to electron
pairing, it does not lead to a coherence peak. In the ground-state of this
insulator all electrons are paired on individual localized eigenfunctions.
When the Fermi-level gets closer to the mobility edge, superconductive
correlations develop between localized pairs. Key features of the predicted
superconductive ground-state are extreme inhomogenity of superconductive
correlations in real space, an unusually large (compared to $T_{c}$)
single-particle excitation gap (spin gap), and pseudogaped regime at
temperatures about $T_{c}$. All these unusual features are due to the
fractal nature of localized eigenstates near the mobility edge.

We are grateful to B. L. Altshuler, T. I. Baturina, E. Cuevas, A. M.
Finkelstein, V. F. Gantmakher, A. S. Ioselevich, A. Millis, A. D. Mirlin, M.
Mueller, Z. Ovadyahu, V. V. Ryazanov, D. Shahar, A. Silva and M. A.
Skvortsov for useful discussions. This research was supported by NSF grants
DMR 0210575 and ECS-0608842, by NATO CLG grant 979979, by RFBR grants
04-02-16348, 04-02-08159 and by the program "Quantum Macrophysics" of RAS.
E.Y. was supported by Alfred P. Sloan Research Fellowship and NSF grant DMR
0547769.


\begin{thebibliography}{99}
\bibitem{absense} which is absent in our case because of the frequency 
dependence in Eq.(\ref{MM}) that eliminates the logarithmic corrections
(present in usual BCS problem due to virtual transitions involving
higher-energy states).
\bibitem{fraction} This estimate comes from the following arguments:
eigenstates that participate in the formation of a pairing amplitude $F(%
\mathbf{r})$ are those which have energies within the range $\Omega \sim T_c$
around the Fermi energy. Enchancement of matrix elements $M_{jk}$
demonstrated by Eq.(\ref{MM}) comes from the fact that the corresponding
eigenfunctions are all concentrated within the same fraction,  $%
\delta^\Omega = \delta \mathcal{V}(\Omega)/\mathcal{V}$, of the total
volume. Therefore $\delta^\Omega$ can be estimated as \textit{inverse} of
that enchancement factor, $\delta^\Omega \sim (\Omega/E_0)^\gamma$. One more
comment is in order here: implicitely, we assumed that all relevant
eigenstates within $\Omega \sim T_c$ range are characterized by nearly the
same $\delta_L$. This assumption is good as long as $(E_m-E_F)/E_0 \equiv
\tau \gg T_c$. Note that this inequility is consistent with "critical"
behaviur realized at $\delta_L \ll T_c$, since $\delta_L \sim E_0\tau^{3\nu}$%
, and $\nu \geq 1$.
\end{thebibliography}
\end{document}